\documentclass[12pt]{article}

\usepackage{helvet}
\usepackage[margin=1.0in]{geometry}
\usepackage{titling}
\usepackage{setspace}
\usepackage[numbers]{natbib}
\usepackage{fancyhdr}
\usepackage[hidelinks]{hyperref}
\usepackage{authblk}
\usepackage{IEEEtrantools}

\renewcommand\Affilfont{\itshape\small}
\makeatletter
\renewcommand\AB@affilsepx{\quad\protect\Affilfont} 
\makeatother

\makeatletter

\renewcommand\maketitle{
{\raggedright 
\begin{center}
  {\setstretch{1.2}\Large \bfseries \@title\par}
  \vspace{2ex}
  {\setstretch{1.2}\normalsize \bfseries Response to RFI: Implementation and Changes to Science Policy Document (SPD)-41: Science Information Policy\par}
  \vspace{2ex}
{ \@author}\\[2ex]
\@date\\[3ex]
\end{center}}} 
\makeatother

\title{Comments on SPD-41 software licensing requirements}
\author[1]{John D. Haiducek, john.haiducek@nrl.navy.mil}
\affil[1]{U.S. Naval Research Laboratory}
\date{18 February 2022}
\begin{document}

\maketitle

\begin{abstract}
  The proposed changes to Science Missions Directorate (SMD) Policy Document 41 (SPD-41) include requirements that SMD-funded software should be released using a permissive software license and following best practices from the open source software community. This is a welcome change that will lead to greater acceptance and adoption of SMD-funded software. However, ambiguities exist in the policy text around what licenses will be allowed. This could lead to problems in the solicitation process and friction with the open source software community. Moreover, some additional clarifications are warranted around what exceptions might be allowed to the software licensing policy.
\end{abstract}

Software licensing is currently a subject of friction between NASA, the scientific community, and the open source software community. This is largely the result of scientists both inside and outside of NASA lacking knowledge around open source licensing. Many scientists understand terms like ``open source,'' ``free software,'' and ``permissive license'' to simply mean that the source code is available in some form. They are unaware of the rigorous definitions applied by the Open Source Initiative (OSI) and Free Software Foundation (FSF) that determine what licenses are considered acceptable by the Free/Open Source Software (FOSS) community. Some scientists are resistant to the idea of releasing their work as open source, wishing to hold their work close as a means of maintaining a competitive edge or protecting their reputation. The proposed changes to SPD-41 will help alleviate some of this friction, but ambiguities in the text will lead to continued misunderstandings. This could create problems in SMD solicitations and barriers to adoption of software outside NASA.

\subsection*{How the proposed changes may impact the research and related activities of SMD communities}

The primary impact of the proposed software licensing requirements in SPD-41 will be to encourage researchers to use permissive open source software licenses. This will benefit both NASA and the research community by making SMD-funded software more widely available. Releasing software openly makes it possible for outside groups to use and contribute to the software. Releasing it under a permissive license lowers the barriers to adoption, maximizing the number of people who are able to benefit from the work and contribute improvements to the software.

\subsubsection*{Lack of clarity in the SPD-41 language}

Paragraph III.C.a. of the proposed revision to SPD-41 states that ``SMD-funded software should be released under a permissive license that has broad acceptance in the community.'' Similarly, Paragraph V.B. requires that ``research software developed using SMD funding...should be released as open source.'' The terms ``open source'' and ``permissive license'' are defined in Appendix B. \textbf{The provided definitions are superficially similar to the definitions of ``open source'' and ``free software'' provided by the OSI \citep{OpenSourceDefinition} and FSF \citep{WhatFreeSoftware}, but are considerably less clear.}

The definition of ``open source software'' provided in Appendix B. mentions a few of the key FOSS principles, but others, such as the right to understand how a program works and use it for any purpose, are omitted. \textbf{The OSI and FSF definitions are mentioned, but are characterized as informational rather than binding}, with the words ``OSS is often distributed under licenses that comply with the definition of `Open source' provided by the Open Source Initiative or meet the definition of `Free Software' provided by the Free Software Foundation.'' Thus it is unclear what role, if any, the OSI and FSF definitions will play in implementation of SPD-41. This ambiguity could be used to justify claims that a wide variety of license restrictions are consistent with SPD-41, including restrictions that would render the license non-free or non-open under the OSI and FSF definitions.

The OSI defines a permissive license as being open source by definition \citep{FrequentlyAnsweredQuestions}, and thus it is implied that a permissive license complies with all the terms in the OSI definition of Open Source \citep{OpenSourceDefinition}. The same interpretation does not follow from the definition of ``permissive license'' in SPD-41. Like the definition of ``open source,'' it summarizes a few FOSS principles but leaves much to interpretation. The OSI and FSF definitions make it explicit that the word ``free'' pertains to specific aspects of user freedom and not, for instance, to the sale price of the software. SPD-41, on the other hand, \textbf{does not specify what rights and privileges are implied by ``free use,'' ``free modification,'' and ``free redistribution.''}

Furthermore, it is not even clear whether the definitions of ``permissive license'' and ``open source software'' in Appendix B. are intended to be a binding part of the policy, or merely information provided to aid the reader in understanding. As a result, a person reading the proposed policy text might simply read the words ``permissive license'' and interpret language in a manner suitable to their interests, ignoring even the loose definitions provided in the appendix.

From reading the proposed SPD-41 text, it is also \textbf{not clear whether SMD will fund modification of existing software that does not use a permissive license}. A strict interpretation of paragraph III.C.a. would suggest that this will not be permitted. However, paragraph II.D.a. states that information subject to ``patent or intellectual property law'' may be eligible for exceptions under SPD-41. It is conceivable that SMD might use this clause as grounds to allow exceptions to SPD-41 for any work that proposes to modify existing software. Or an exception might only be made for cases involving modification of software for which the SMD-funded developers are not authorized to change the license.

Finally, paragraph II.B.c. of SPD-41 provides definition of ``software'' that includes the unusual constraint that the code provide ``some degree of scientific utility or produce a scientific result or service'' in order to be considered software. This could lead to confusion as to the applicability of SPD-41 to software that is not directly related to science.

\subsubsection*{Barriers to adoption}

The proposed policy changes \textbf{should be expected to lower barriers adoption} for any software whose development is funded by SMD. The requirement to release software under an permissive license that has ``broad acceptance in the community'' will undoubtedly make the software more available to the broader community. However, \textbf{differences between the definitions in SPD-41 Appendix B and the FOSS definitions provided by OSI and FSF could lead to continued friction} between NASA and the open source software community.

In the past, subtle problems with license language have led to licenses being rejected by OSI and/or FSF, a notable example being the NASA Open Source Agreement (NOSA), which was approved by OSI as an open source license \citep{OpenSourceLicenses} but rejected by FSF as being non-free \citep{VariousLicensesComments}. The impact of this goes beyond criticism on a web page, because compliance with the OSI and FSF definitions is used to determine whether a software is eligible for inclusion in software repositories and operating systems. In the past, some NASA-developed software has been rejected for inclusion in Linux distributions because of license language, despite community interest in having the software included \citep{ReviewRequestCdf,LicensingMainFedora}. Issues like this have been identified both inside and outside NASA as barriers to adoption of NASA-funded software \citep{NationalAcademiesofSciencesEngineeringandMedicine2018OpenSourceSoftware,Beyer2018NoNOSAYes}.

Given the history of friction between NASA and the open source software community, considerable care is warranted in addressing this. The proposed changes to SPD-41 do constitute an improvement over previous policies and practices. However, \textbf{ambiguities in the policy text mean that some licenses may be allowed under SPD-41 that do not pass muster with the FOSS community}. Although this could be addressed by revising the definitions in Appendix B (and making it clear that they are binding), a better approach would be to simply \textbf{require the use of software licenses that comply with both the OSI and FSF definitions}, rather than attempting to create a new definition. Furthermore, ``broad acceptance in the community'' should be clarified to mean acceptance by both OSI and FSF. This would ensure the broadest possible acceptance of SMD-funded software within the open source software community.

\subsubsection*{Fairness in solicitations}

As a result of the ambiguity in various parts of the proposed text, the software licensing requirements in SPD-41 will likely be interpreted differently by some scientists than by others. This could lead to problems of fairness in the solicitation process. Some proposal writers might interpret the policy strictly, while others might interpret it more loosely.

Those who interpret the policy more strictly than NASA intends might needlessly restrict the scope of their proposed work. For instance, they might choose not to propose certain work because it would require modifying existing software that is distributed under a non-permissive license. Or they might propose to create new software to support their work rather than modify existing software that uses a non-permissive license. Thus proposal teams that strictly interpret the software licensing provisions in SPD-41 could find themselves at a disadvantage relative to their competition.

Those who interpret the policy more loosely than NASA intends would encounter a different problem. They might expend effort writing a proposal that the believe to be in compliance, but which uses a license that is non-compliant in some way. This could lead to proposals being rejected solely due to software licensing issues or to late-stage modifications to comply with SPD-41 policy, which could be avoided by having clearer language in SPD-41.

Another issue that can arise in solicitations is that proposals may provide insufficient information to determine whether the proposed work will comply with SPD-41. For instance, a proposal might commit generally to using a permissive license, without specifying what software license will be used. In such cases, the proposal team might intend to use a non-FOSS or non-permissive license which they mistakenly believe will be in compliance with SPD-41. If the proposal did not specify what license will be used, SMD would not be aware of the problem until later in the process. To prevent this, SMD should consider requiring \textbf{all proposals provide a list of all software products the proposers anticipate creating or modifying, and what software license will apply to each.}

\subsubsection*{Applicability to third-party software}

It is possible that the requirement to distribute SMD-funded software under a permissive license might prevent SMD from funding work that involves modifications to existing software. Many established scientific codes are treated as proprietary and distributed with a restrictive license or with no license. Others are distributed under copyleft (i.e. non-permissive) open source licenses. While the copyright holders are free to change the licenses of these codes in order to comply with the provisions of SPD-41, others do not have this ability. As discussed above, paragraph II.D.a. of SPD-41 seems to allow a way for SMD to make exceptions to allow SMD to fund modifications of third party software that is not in compliance with paragraph III.C.a. If such exceptions are not allowed, the software licensing requirements could significantly restrict the software development activities that SMD is able to fund.

\subsubsection*{Effects on solicitation responses}

Lack of comfort or familiarity with FOSS licenses could cause some scientists to become hesitant in proposing to SMD solicitations. Scientists who are unfamiliar with FOSS licenses might lack confidence in choosing a license for their project. Others might prefer not to use FOSS licenses due to a desire  to maintain a competitive advantage relative to their peers, to prevent commercial users profiting off their work, or to protect themselves from reputational harm resulting from misuse of their software. Others might be comfortable with FOSS licenses but be confused by SMD policy language around licensing. Clearer policy language and better education and outreach around software licensing would help address these problems and maintain robust responses to SMD solicitations.

\subsection*{Support, training, and guidance needed to support successful implementation of the policy}

\textbf{Relatively few scientists are well versed in the subtleties of software licensing.} As a result, SMD should plan to provide guidance to help scientists with choosing software licenses that will be compliant with SPD-41. Even those who are knowledgeable in this area will require additional guidance on how SPD-41 will be implemented, especially if ambiguities in the policy are not addressed in a future revision.

\subsubsection*{Education on software licensing}

Scientists proposing to SMD may require additional education about software licensing in order to comply with the proposed SPD-41 changes. This includes training on the following topics:

\begin{itemize}
\item Considerations for choosing a software license
\item How to determine whether a license complies with SPD-41
\item What existing licenses are in compliance with SPD-41
\item How developers can protect their own interests while distributing their work under a permissive license
\item How the use of a permissive license can benefit software developers, SMD, and the wider community
\end{itemize}

\subsubsection*{Guidance on how SPD-41 will be implemented}

As noted earlier, there are several aspects of software licensing policy that are poorly defined in SPD-41. \textbf{Ideally, some of this ambiguity should be removed by revising the policy itself.} If the current text becomes policy, SMD stakeholders will need guidance in the following areas to address aspects of implementation that are not fully addressed in the SPD-41 text:

\begin{itemize}
\item How SMD will determine whether SPD-41 applies to a particular piece of software
\item How SMD will determine whether a license is ``permissive''
\item How SMD will determine whether a license has ``broad acceptance in the community''
\item Under what circumstances, if any, SMD will make exceptions to fund modifications to existing software that does not meet the proposed software licensing requirements
\item What information about software licensing needs to be provided in proposals to SMD
\end{itemize}

\bstctlcite{IEEEexample:BSTcontrol}

{\footnotesize
  \bibliography{IEEEtranBSTCTL,sources}}

\end{document}